%% file: main.tex
\documentclass[sigconf, authorversion]{acmart}

\usepackage{booktabs,tabularx}
\usepackage{natbib}
\usepackage{balance}
\usepackage{hyperref}
\usepackage{cleveref}
\newcommand{\multicell}[2][t]{\begin{tabular}[#1]{@{}l@{}}#2\end{tabular}}

\begin{document}

\copyrightyear{2018}
\acmYear{2018}
\setcopyright{acmlicensed}
\acmConference[SEEM'18]{IEEE/ACM International Workshop on Software Engineering Education for Millennials }{May 27-June 3 2018}{Gothenburg, Sweden}
\acmBooktitle{SEEM'18: IEEE/ACM International Workshop on Software Engineering Education for Millennials , May 27-June 3 2018, Gothenburg, Sweden}
\acmPrice{15.00}
\acmDOI{10.1145/3194779.3194784}
\acmISBN{978-1-4503-5750-0/18/05}

\title[Scrum2Kanban]{Scrum2Kanban: Integrating Kanban and Scrum in a University Software Engineering Capstone Course}

\author{Christoph Matthies}
\affiliation{%
  \institution{Hasso Plattner Institute, University of Potsdam}
  \streetaddress{August-Bebel-Str. 88}
  \city{Potsdam} 
  \state{Germany} 
  \postcode{14182}
}
\email{christoph.matthies@hpi.de}


\begin{abstract}
Using university capstone courses to teach agile software development methodologies has become commonplace, as agile methods have gained support in professional software development.
This usually means students are introduced to and work with the currently most popular agile methodology: Scrum.
However, as the agile methods employed in the industry change and are adapted to different contexts, university courses must follow suit.
A prime example of this is the Kanban method, which has recently gathered attention in the industry.
In this paper, we describe a capstone course design, which adds the hands-on learning of the lean principles advocated by Kanban into a capstone project run with Scrum. This both ensures that students are aware of recent process frameworks and ideas as well as gain a more thorough overview of how agile methods can be employed in practice.
We describe the details of the course and analyze the participating students' perceptions as well as our observations. We analyze the development artifacts, created by students during the course in respect to the two different development methodologies.
We further present a summary of the lessons learned as well as recommendations for future similar courses. The survey conducted at the end of the course revealed an overwhelmingly positive attitude of students towards the integration of Kanban into the course.
\end{abstract}

%
%



\maketitle

\input{CONTENT}

\let\oldbibitem\bibitem
\def\bibitem{\vfill\oldbibitem}

\newpage
\balance

\bibliographystyle{ACM-Reference-Format}
\bibliography{library}

\end{document}

%% file: CONTENT.tex
\section{Introduction}
Teaching software development methods and processes in capstone courses using hands-on projects is a widely adopted practice in universities~\cite{Mahnic2015}.
These projects strengthen comprehension and allow students to apply their theoretical process knowledge in real-world software projects~\cite{Paasivaara2017, Paasivaara2013, Kropp2013}.
The 2013 CS curriculum guidelines jointly published by ACM and IEEE state that the opportunity for students to iteratively work through a development cycle, assessing outcomes and applying their gained knowledge positively impacts learning success~\cite{JointTask2013}.
Teaching iterative agile and lean software development methods has thus, in addition to fulfilling industry needs, become an important issue when designing software engineering university courses~\cite{Mahnic2015a}.
Most of these courses focus on the Scrum methodology~\cite{Paasivaara2017, Persson2011, Rico2009}, as it is currently the most popular process employed in industry settings~\cite{stateofagile11} and has become mainstream~\cite{Kropp2014a}.
Much of the innovation in the domain of software development processes happens in industry, driven by a need for practices that increase software quality and enable development teams to work together more effectively.
Therefore, the software engineering curricula must respond to changes in how agile and lean software development methods evolve in practice.

While in most organizations Scrum is still the most employed development methodology, lean approaches such as Kanban are on the rise.
The latest ``State of Agile Report'' by VersionOne, published in 2017, identified Scrum and Scrum-XP-Hybrids as the most common agile methodologies, used by 68\% of respondents.
These methods were followed directly by custom hybrids (8\%), Scrumban (8\%) and Kanban (5\%)~\cite{stateofagile11}.
Other recent surveys confirm this trend.
Komus et al. state that in their survey of more than 1000 respondents, Scrum is the most often employed agile methodology (85\% of respondents), followed by Kanban and Lean~\cite{Komus2017}.
The Scrum Alliance reports that 89\% of respondents to their ``State of Scrum'' survey named specifically Scrum as the agile approach or at least one of the agile approaches used in their organization~\cite{ScrumAlliance2017}.
Again, Kanban was the second most commonly named software development approach.

In order for university software engineering courses to stay relevant and up to date, Kanban and lean software development approaches need to be integrated into the curriculum.
To this end, we present an undergraduate project course design, based on a modified version of Scrum, which introduces Kanban towards the end of the instruction period and allows students to gain a better overview of agile methods, in line with recent research~\cite{Mahnic2015a}.
We evaluate our approach both with surveys as well as analysis of the produced development artifacts, e.g. commits and user stories/tickets.


The following research questions guide our work:
\begin{enumerate}
\renewcommand{\labelenumi}{\textbf{RQ\arabic{enumi}}}
\item \label{r1} How can the Kanban software development methodology be integrated into an agile hands-on undergraduate university software engineering course?
\item \label{r2} What are students' perceptions of Kanban practices and the proposed teaching approach?
\item \label{r3} Are students' perceptions reflected in the development artifacts that were produced while employing Kanban?
\item \label{r4} What influence does using Kanban have on the employed workflow during the course?
\end{enumerate}


The rest of the paper is structured as follows: Section~\ref{sec:scrum} briefly introduces the agile methods Scrum and Kanban as well as possible hybrids employed in the industry. Section~\ref{sec:course_design} describes the course design and details the development process that is followed during the course. The following section, Section~\ref{sec:evaluation}, describes how we evaluated the chosen teaching approach using surveys and analysis of development data. Section~\ref{sec:related_work} details related work in the research areas of teaching agile capstone courses and evaluating development processes. The last section, Section~\ref{sec:conclusion} concludes and summarizes our findings.

\section{Scrum, Kanban and Hybrids} 
\label{sec:scrum}
Scrum as well as Kanban are agile software development methods.
They focus on collaborative teamwork and explicitly acknowledge the importance of self-organization and empowerment of development teams~\cite{Kniberg2009}.
Scrum is the more well-known of the two approaches and relies on an iterative and incremental development cycle, with the goal of producing an increment of potentially shippable software at the end of each ``Sprint'' iteration.
It focuses on managing projects by introducing defined roles (e.g. Product Owner), meetings (e.g. Sprint Retrospective) as well as organizational and process artifacts~\cite{Schwaber2004}.
The Kanban development method is inspired by the Toyota Production System~\cite{Ohno1978} and the principles of Lean manufacturing~\cite{james1991machine}.
In contrast to Scrum, it is less descriptive and focuses on visualizing workflows using Kanban boards, limiting the work in progress and ensuring that work flow as fast as possible through the system by removing bottlenecks~\cite{Kniberg2009}.

Due to their different focuses, Scrum practices and Kanban principles can be combined, resulting in a development process often referred to as \emph{Scrumban}~\cite{Ladas:2009}.
In a study on the adoption of Scrum in enterprises Kapitsaki et al. identified the need for more studies that follow Scrum's adoption and its emerging variants, especially in combination with Kanban~\cite{Kapitsaki2014}.
In a 2015 literature review on the topic of process models in practice Theocharis et al. state that hybrid approaches, including ``traditional-agile'' as well as ``agile-agile'' hybrids represent a common model of use~\cite{Theocharis2015}.
They further state that in their set of 22 selected papers on process usage, only five provided quantitative data and call for more instruments for process tailoring and combination.
A recent online study, performed in 2018 by Kuhrmann et al. on the characteristics of hybrid development approaches confirmed these findings.
The authors state that among the 69 surveyed practitioners across Europe, hybrid development methods were widely used, regardless of company size or industry domain and were applied even in when company-wide policies for process use were present~\cite{Kuhrmann2018}.
As such, teaching Kanban and its possibilities and combinations is vital for educating the next generation of well-rounded software engineers.

\section{Course Design}
\label{sec:course_design}


In order to answer our first research question (\emph{RQ\ref{r1}}) on how Kanban can be integrated in a university software engineering course, we describe an undergraduate capstone course design, which was developed over the last 4 years and was most recently taught in the winter semester of 2017/18.
Its main focus is collaboration and self-organization in teams and in a team-of-teams using agile methods as well as modern software development best practices.
The course emphasizes hands-on learning in a simulated real-world scenario, its main learning targets are summarized in Table~\ref{table:learning_targets}.
All students of the course, split into teams, jointly develop a single software system under an open-source license, hosted on the collaboration platform GitHub\footnote{\url{https://github.com/}}.
Work in the project is accompanied by regular lectures as well as support by tutors (junior research assistants), who are present during students' team meetings.
All materials and slides of the latest iteration of the course are available online\footnote{\url{https://hpi.de/plattner/teaching/winter-term-201718/softwaretechnik-ii.html}}.
As the course is recommended for students in the last semester of undergraduate studies, students have already attended lectures on the fundamentals of software engineering.
The Scrum method is employed in four sprint iterations in the beginning of the course, as its more descriptive nature and requirements for structure lend themselves to introducing agile concepts~\cite{Mahnic2015a}.
After students have gained experience with the employed technologies and have become familiar with their teams, the more dynamic Kanban method is employed.
This shift in process introduces additional challenges and must be accompanied with tutoring of student teams as well extensive lectures.

\begin{table}[tb]
    \begin{tabularx}{\columnwidth}{|l|X|}
        \hline
        \textbf{\#} & \textbf{Learning target}\\ \hline
        1 & Experience with Kanban as well as Scrum and all of its artifacts and meetings\\ \hline
        2 & Knowledge of how to scale Scrum over multiple collaborating teams\\ \hline
        3 & Ability to use BDD and TDD where appropriate\\ \hline
        4 & Confidence with branching and merging in a source code management (SCM) system\\ \hline
        5 & Experienced the value of continuous integration (CI)\\ \hline
        6 & Learned to critically self-assess one's own role in a team\\ \hline
    \end{tabularx}
    \caption{Overview of the main learning targets of the software engineering II course.}
    \label{table:learning_targets}
\end{table}

\subsection{Course Schedule}
The course described here is typically run in the winter semester with a length of 15 weeks.
It features both traditional lectures, introducing students to agile methods and best practices, as well as stretches of project work and separate exercises, Table~\ref{table:schedule} lists a typical schedule.
Lectures (e.g. on Scrum and git) are more frequent towards the beginning of the course in order to prepare students for the hands-on project.
Later lectures focus on tips and tricks, guest lectures from industry practitioners and introduce the Kanban method.
During the project part of the course, students jointly develop a web-application using the Ruby on Rails framework\footnote{\url{http://rubyonrails.org/}} in four Scrum sprints, followed by a week in which the Kanban method is practiced.

\begin{table}[tb]
    \begin{tabularx}{\columnwidth}{|l|l|X|}
        \hline
        \textbf{Week} & \textbf{Project work} & \textbf{Lectures}\\ \hline
        1 & & \multicell{\textbullet~Course introduction\\\textbullet~Technology tutorial}\\ \hline
        2-4 & \multicell{Introduction\\exercise} & \multicell{\textbullet~Scrum methodology\\\textbullet~LEGO exercise\\\textbullet~Software testing\\\textbullet~Git version control system} \\ \hline
        5-6 & Sprint 1 & \\ \hline
        7-8 & Sprint 2 & \multicell{\textbullet~Code Reviews\\\textbullet~Deployment}\\ \hline
        9-11 & \multicell{Sprint 3 \&\\Interm. presentation}& \multicell{\textbullet~Scrum Tips\\\textbullet~Guest Lecture}\\ \hline
        12-13 & Sprint 4 & \multicell{\textbullet~Guest lecture\\\textbullet~Lean Software and Kanban}\\ \hline
        14 & Kanban Week & \\ \hline
        15 & Final Presentation & \textbullet~Summary, exam preparation\\ \hline
    \end{tabularx}
    \caption{Course schedule. Project work is performed by students in their teams or individually on their own schedule.}
    \label{table:schedule}
\end{table}

In order to allow students to get started with hands-on project work as fast as possible, two practical exercises are employed at the beginning of the course: A LEGO Scrum exercise and an automated programming tutorial.

After a lecture introducing the basics of Scrum, the roles, meetings and artifacts involved, a 90 minute ``fast-track'' Scrum simulation exercise is set.
This sort of exercise, where LEGO is built in sprints of a few minutes each, has been shown to engage students, simulating the phases and difficulties of real software projects and advances student learning~\cite{Paasivaara2014}.

Furthermore, an interactive programming exercise aims to familiarize students with the web application framework used in the following development project.
The exercise, which students have 3 weeks to complete, involves an automated Continuous Integration (CI) server, which analyzes students' progress and assigns them new tasks through GitHub~\cite{Matthies2017}.
Students are thus introduced not only to the programming framework, but also all the other tools necessary, such as dependency managers, testing frameworks, editors and CI servers.

\subsection{Scrum Development Process}

\begin{figure}[htb]
\centering
    \includegraphics[width=\columnwidth]{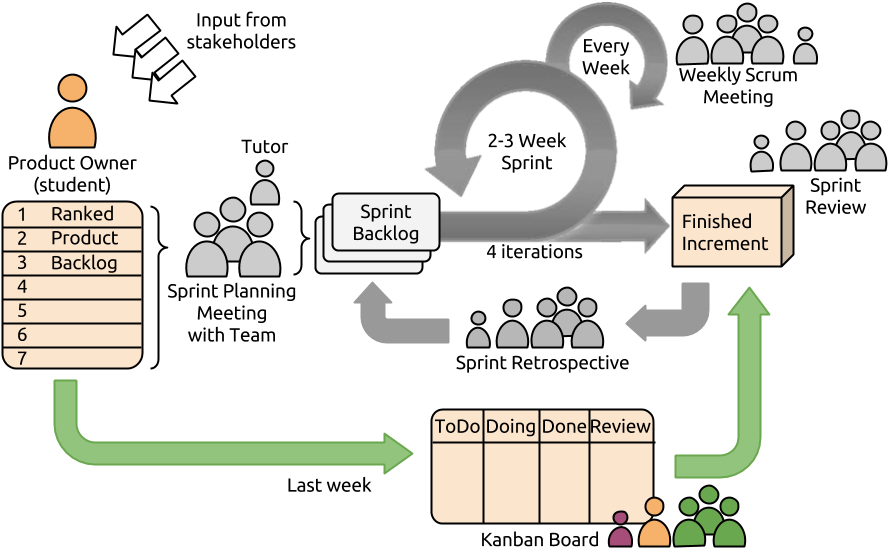}
    \caption{Overview of the modified Scrum process (top, grey) and the integration of Kanban (bottom, colored) used in our software engineering course.}
    \label{fig:process}
\end{figure}

During the four sprints of the project, a version of the Scrum process, adapted to the time constraints of students is employed, see Figure~\ref{fig:process}.
Participants form self-teams of 5 to 8 members, each with their own Product Owner (PO) and a Scrum Master (SM), roles which are also performed by students.
All other students are developers.
During each week of the course, students are expected to spend 8 hours of work time, including lectures and team meetings.
Due to these time constraints, daily Scrum stand-up meetings are replaced by weekly versions.

For every sprint, a Scrum planning, sprint review, retrospective as well as a weekly stand-up meeting is organized by the students.
These meetings are held in the presence of a tutor, who is able to answer questions and provide guidance and advice, if needed.
In order to elicit the initial requirements for the system to be developed during the project, the POs meet with a member of the teaching staff, the customer.
Guided by an instructor, taking the role of ``chief PO'', the product backlog is filled with user stories.
The Product Owners then present the product vision to their teams and the first Sprint is started.
During the middle of the project, the POs hold an intermediate presentation, detailing the current status of the software to the entire course as well as external stakeholders that will use the software.

\subsection{Kanban Development Process}
Kanban is considered, relatively speaking, less \emph{prescriptive} and more \emph{adaptive} than Scrum, meaning there are fewer rules to follow~\cite{Kniberg2009}.
This means Kanban introduction and teaching is better suited when students have already acquired an understanding of agile concepts and can interpret and adapt the process according to their needs and context.
Therefore, in order to improve the learning of Kanban, it is advisable to introduce the method after the students have already gained experience with Scrum during the project~\cite{Mahnic2015a}.
In fact, Kanban users are expected to experiment  with and continually improve their process in order to customize it to their environment, a task that is beyond simple application of learned methods and practices~\cite{Nikitina2012, Mahnic2015a}.
While it can be argued that systems should be stable, i.e. a development process in a team that has a long history, before a new process is introduced, Kanban can also provide mechanisms to achieve the needed stabilization by focusing on throughput~\cite{reddy2015scrumban}.

Towards the end of the course, after students have become familiar with Scrum and their teams, i.e. the \emph{norming} or \emph{performing} stages of group development~\cite{Bonebright2010}, the ideas of Lean Software development and Kanban are introduced in a lecture, see Table~\ref{table:schedule}.
Kanban boards and the concepts of visualizing workflow and limiting work in progress (WIP) through WIP limits are presented in detail.
students discuss the advantages and drawbacks of the method, compared to Scrum and other processes they previously experienced.
The course attendees are then encouraged to try out these new ideas in their project by employing Kanban for the last week to 10 days of the project, instead of a last Scrum sprint, see Figure~\ref{fig:process}.
Kanban introduction is furthermore most suited towards the end of the course, due to the natural way of how work is performed near the deadline of the project.
As the course end and the date of the final presentation are known in advance, students' focus in the last few project weeks traditionally shifts from a ``feature implementation'' to a ``bug fixing and polishing'' mode, in order to deliver a usable project that can be presented.
This work mode entails dynamic and often changing new requirements handed to the developers by the Product Owners, as bugs are fixed, new ones are uncovered, and the software is made ready for presentation.
Therefore, the concept of a Sprint iteration, at the beginning of which the set of user stories to work on are decided and which is shielded from interference for the duration of the sprint, is not ideal.
By employing Kanban, students are allowed to finish up their project using an agile methodology that can adapt to their requirements.

\section{Evaluation}
\label{sec:evaluation}
To understand students' perceptions of Kanban practices and to analyse whether the introduction of a new methodology successfully changed the workflow, we performed two types of evaluations: A survey at the very end of the course (before grades were announced) and an comparative analysis of the development artifacts produced during Kanban and Scrum project work.

\subsection{Survey}
We conducted a voluntary, anonymous online survey among all participants of the 2016/17 installment of the course.
The survey of 11 questions focused on students' perceptions of Kanban practices, the biggest advantages and drawbacks and adaptations to the workflow that were performed.
All questions of the survey are listed in Table~\ref{table:survey}.

\begin{table}[htb]
    \begin{tabularx}{\columnwidth}{|l|l|X|}
        \hline
        \textbf{\#} & \textbf{Type} & \textbf{Question Text}\\ \hline
        1a & \multicell{5-point\\scale} & Was the Kanban week at project end more useful and productive then a last week of Scrum? \\ \hline
        1b & \multicell{5-point\\scale} & Did the Kanban week at project end offer a good insight into other agile development methods? \\ \hline
        2 & \multicell{5-point\\scale} & Did you have to adapt your workflow for the Kanban week?\\ \hline
        3a & free text & What were the biggest advantages of using Kanban in your team?\\ \hline
        3a & free text & What were the biggest disadvantages of using Kanban in your team?\\ \hline
        4 & choice & How did user stories change from using Scrum to Kanban?\\ \hline
        5a & free text & In the next iteration of the course, how long should Kanban be used?\\ \hline
        5b & free text & In the next iteration of the course, when should Kanban be employed (e.g. in the beginning)?\\ \hline
        6 & \multicell{5-point\\scale} & Would you recommend using Kanban to the participants of next year's course?\\ \hline
        7 & \multicell{5-point\\scale} & Should there have been an additional lecture concerning Kanban?\\ \hline
        8 & free text & Own feedback, e.g. on best practices to be followed for the next course?\\ \hline
    \end{tabularx}
    \caption{Questions of the student survey performed at the end of the software engineering course.}
    \label{table:survey}
\end{table}

Of the 22 attendees of the course, 18 (17 men, one woman) participated in the survey.
In addition to questions that could be answered on a 5-point Likert scale (1: strong no, 2: no, 3: neutral, 4: yes, 5: strong yes), the survey included free text questions as well as a question featuring multiple choice options.
They survey could be submitted without all questions being answered.
However, all Likert-scale questions were answered by all participants.
An overview of the answers to these questions is presented in Table~\ref{table:analysis}.

\begin{table}[tb]
    \begin{tabularx}{\columnwidth}{|l|X|r|r|r|r|r|}
        \hline
        \textbf{\#} & \textbf{\multicell{Question\\Topic}} & \textbf{Mean} & \textbf{\multicell{Std.\\Dev.}} & \textbf{\multicell{10\%\\Trim.\\Mean}} & \textbf{\multicell{Med-\\ian}} & \textbf{Range} \\ \hline
        1a & Kanban week preferred over another Scrum week? & 4.08 & 1.38 & 4.30 & 5.00 & 4.00\\ \hline
        1b & Kanban offered insight into agile methods? & 3.75 & 1.29 & 3.90 & 4.00 & 4.00\\ \hline
        2 & Was the workflow adapted? & 3.83 & 1.11 & 4.00 & 4.00 & 4.00\\ \hline
        6 & Recommended for next year? & 4.33 & 0.98 & 4.50 & 5.00 & 3.00\\ \hline
        7 & Additional lecture on Kanban? & 3.42 & 1.08 & 3.40 & 4.00 & 3.00\\ \hline
    \end{tabularx}
    \caption{Summarized answers of participants to the 5-point Likert scale questions of the survey. Answer possibilities: 1~(strong~no), 2 (no), 3 (neutral), 4 (yes) 5 (strong yes).}
    \label{table:analysis}
\end{table}

\begin{figure}[htb]
\centering
    \includegraphics[width=\columnwidth]{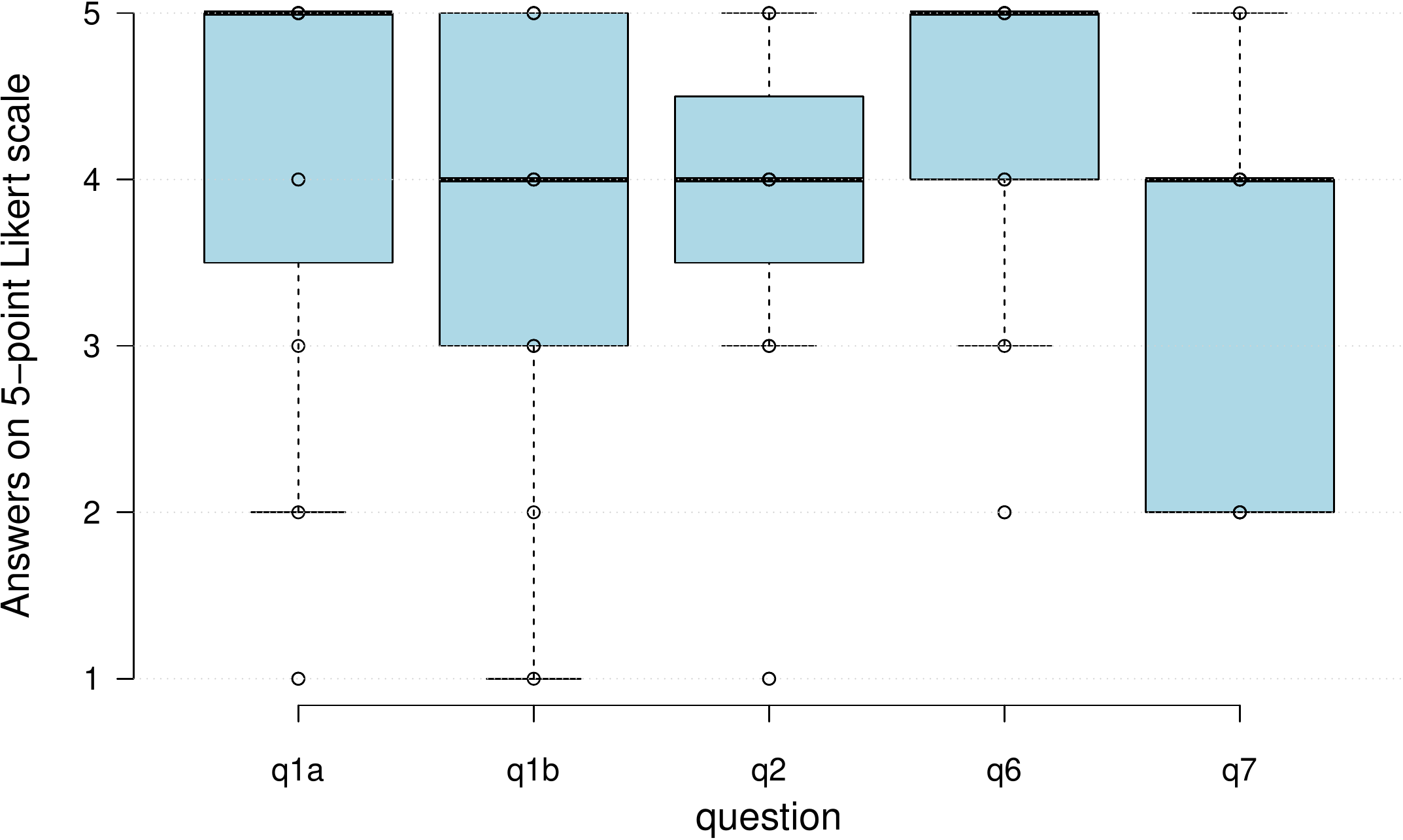}
    \caption{Boxplot of answers to questions 1a, 1b, 2, 6 and~7. Center lines show the medians, box limits indicate the 25th and 75th percentiles, whiskers extend 1.5 times the interquartile range from the 25th and 75th percentiles, outliers are represented by dots.}
    \label{fig:boxplot}
\end{figure}

Overall, students showed positive attitudes towards the inclusion of Kanban in the course. Both questions regarding the use of Kanban (questions 1a, 1b) were answered, on average, positively, with mean values of 4.08  and 3.75 respectively. Students stated that they had adapted their workflows for Kanban (question 2). It is encouraging that students were enthusiastic about including Kanban in the next iteration of the course as well (question 6). The only question, where answer mean as well as 10\% trimmed mean were close not positive, see Figure~\ref{fig:boxplot}, was on the topic of additional lectures (question 7). This is somewhat expected, due the the different learning styles of students, with some enjoying self-study and others more guidance through lectures.

\begin{figure}[htb]
\centering
    \includegraphics[width=\columnwidth]{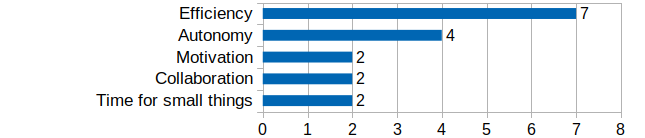}
    \includegraphics[width=\columnwidth]{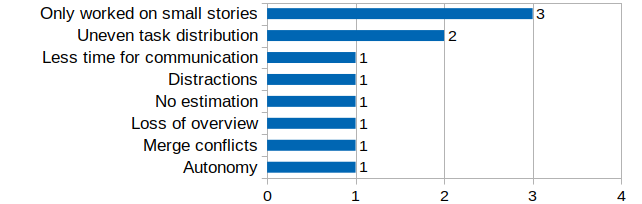}
    \caption{Topics of survey participants' answers to free text questions \#3a and \#3b, concerning the perceived advantages (top, N=11) and drawbacks (bottom, N=9) of using Kanban. An answer could contain multiple topics.}
    \label{fig:q3}
\end{figure}

The free text answers to questions 3a and 3b were manually tagged with the topics they included, the set of topics was iteratively refined after every question was processed.
The results of these two questions are summarized in Figure~\ref{fig:q3}.

Concerning the perceived advantages of applying the Kanban method (question 3a, 3b), students most often mentioned the topics of efficiency and autonomy, concepts close to Lean Software's principles of \emph{eliminating waste} and \emph{empowering the team}~\cite{poppendieck2003lean}.
The main drawbacks of Kanban as described by students in question 3b, were only working on small user stories (in favor of larger ones which were neglected) and uneven task distribution.
Working mostly on small user stories can be seen as a consequence of autonomy, with developers choosing to work on small issues that can be finished quickly, instead of tackling large problems.
Uneven task distribution is an ongoing challenge in self-organizing teams, which might be addressed in future by more thorough coaching of Scrum Masters and tutors to pay attention to this topic.

\begin{figure}[htb]
\centering
    \includegraphics[width=\columnwidth]{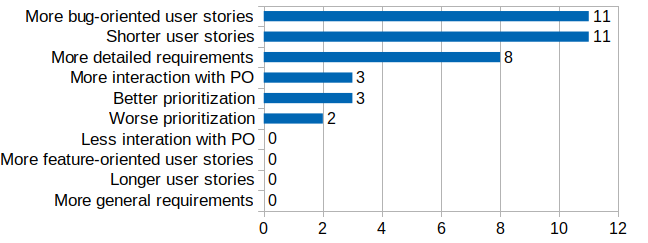}
    \caption{Answers of participants to survey question \#4, concerning attributes of user stories when using Kanban compared to Scrum. N=12.}
    \label{fig:q4}
\end{figure}

The main artifacts that developers and Product Owners use and communicate with, both in Kanban and Scrum, are user stories.
Question 4 of the survey elicited how user stories changed from Scrum to Kanban.
The multiple choice answer possibilities as well as a summary of answers are presented in Figure~\ref{fig:q4}.

As expected for Kanban usage at the end of a project, students classified the user stories that they wrote and worked on as shorter and more bug-oriented than they had previously been using Scrum.
However, students also noted that user stories during Kanban contained more detailed requirements, in terms of acceptance criteria, a good attribute when trying to move tickets through the Kanban board as fast as possible.
Concerning the best time for introducing Kanban in the course, students overwhelmingly agreed with our decision to employ Kanban towards the end of the project for a week, see Table~\ref{table:q5}.

\begin{table}[tb]
    Preferred length of Kanban practice
    \begin{tabular}{|r|r|r|}
        \hline
        \textbf{1 week} & \textbf{1-2 weeks} & \textbf{> 2 weeks} \\ \hline
        6 & 3 & 2 \\ \hline
    \end{tabular}
    
    \vspace{4mm}
    
    Preferred time for Kanban practice
    \begin{tabular}{|r|r|}
        \hline
        \textbf{End of the project} & \textbf{In the middle of the project} \\ \hline
        9 & 1 \\ \hline
    \end{tabular}
    \caption{Summarized answers of students to the free text questions \#5a and \#5b concerning when and how long Kanban should be employed during the course. N=10.}
    \label{table:q5}
\end{table}

The last question on the survey, question 8, asked for feedback on best practices to be followed for the next course iteration in the form of free text. This more reflective answer was only answered by 5 participants. 
However, the answers revealed areas that should be improved in future. Students pointed out that merges were painful and that they lacked a thorough understanding of merge strategies (3 mentions) as well as that instructors should focus on strategies of dealing with errors in others work and should highlight the importance of self-reliance (both 2 mentions). These are areas that will be addressed in future lectures and guest lectures by industry practitioners.


In summary, the survey revealed positive student attitudes towards the teaching approach of a shift from Scrum to Kanban at the end of the project work (\emph{RQ2}), recommending the method to participants of the next year's course.
Students stated that they had adapted their workflow (\emph{RQ4}), which was intended as part of the learning experience and had gained insights into another agile method.
They agreed that Kanban was well suited for the end of the project (rather than the beginning or the middle) and independently identified the advantages of Kanban over Scrum.

\subsection{Development Artifact Analysis}
While surveys allow insights into the perceptions of students, they do not tell the whole story.
It is unclear, solely from survey answers, whether e.g. the perceived change in user story attributes or change in workflow actually happened and how severe the change was.
In order to answer questions such as these, we performed an analysis of the development artifacts produced by course participants during Scrum and Kanban usage (\emph{RQ3}).
All main development activity around producing code and user stories, as tickets in an issue tracker, happened on GitHub, where the project was hosted.
The service features comprehensive APIs\footnote{\url{https://developer.github.com/v3/}}, that allow querying both the issue tracker as well as the git version control system.
For every development iteration, i.e. Sprints 1 to 4 as well as the Kanban week, all created user stories as well as commits were extracted and analyzed.
Using this data, both the assumptions of learning success as well as the perceptions of students could be tested.
Students' perception of shorter user stories while practicing Kanban is reflected in the data.
While mean title length did not change significantly, the mean text body length of user stories was lower in the Kanban week, than in previous sprints, see top of Table~\ref{table:issues}.
The dynamic interactivity with user stories on a Kanban board is reflected in the fact that in the Kanban week only 68.8\% of user stories were created by POs, compared to much higher numbers in the Scrum Sprints, see bottom of Table~\ref{table:issues}.

\begin{table*}[!htbp]
    \begin{tabularx}{\textwidth}{|l|l|l|l|l|l|l|l|l|l|l|l|l|X|}
        \hline
        \multicolumn{2}{|c|}{} & \multicolumn{3}{c|}{\textbf{Title length}} & \multicolumn{3}{c|}{\textbf{Text body length}} & \multicolumn{3}{c|}{\textbf{Comment count}} & \multicolumn{3}{c|}{\textbf{Events count}}\\ \hline
        \textbf{Iteration} & \textbf{\# Issues} & \textbf{Mean} & \textbf{Stdev} & \textbf{Median} & \textbf{Mean} & \textbf{Stdev} & \textbf{Median} & \textbf{Mean} & \textbf{Stdev} & \textbf{Median} & \textbf{Mean} & \textbf{Stdev} & \textbf{Median} \\ \hline
        Sprint 1 & 10 & 45.7 & 12.2 & 46 & 630.4 & 487.3 & 526 & 1.5 & 2.59 & 0 & 21.2 & 3.5  & 21\\ \hline
        Sprint 2 & 17 & 45.9 & 14.7 & 45 & 889.1 & 501.8 & 713 & 2.2 & 2.96 & 1 & 23.5 & 10.5 & 21\\ \hline
        Sprint 3 & 31 & 38.2 & 13.6 & 35 & 804.7 & 405.9 & 768 & 1.5 & 1.86 & 1 & 18.2 & 9.5  & 17\\ \hline
        Sprint 4 & 31 & 48.1 & 17.5 & 44 & 617.5 & 513.1 & 504 & 0.4 & 0.92 & 0 & 13.6 & 5.4  & 12\\ \hline
        Kanban & 32 & 34.8 & 17.2 & 30 & 315.7 & 367.4 & 221.5 & 1.2 & 1.77 & 1 & 7.7  & 3.6  & 7 \\ \hline
    \end{tabularx}
    
    \vspace{4mm}
    
    \centering
    \begin{tabular}{|l|l|l|l|l|}
        \hline
        \textbf{Iteration} & \textbf{Opened by PO} & \textbf{Closed by PO} & \textbf{Same person open/close} & \textbf{Unique assignees count}\\ \hline
        Sprint 1 &  100\% & 60.0\% & 50.0\% &  9 \\ \hline
        Sprint 2 &  100\% & 29.4\% &  5.9\% & 12 \\ \hline
        Sprint 3 & 87.1\% & 12.3\% & 12.9\% & 18 \\ \hline
        Sprint 4 & 90.3\% & 41.9\% & 29.0\% & 15 \\ \hline
        Kanban   & 68.8\% &  6.3\% &  9.4\% & 16 \\ \hline
    \end{tabular}
    \caption{Summary of issues, i.e. user stories, by iteration. Events are all interactions with an issue, e.g. \emph{labeled}, or \emph{assigned}, except commenting on an issue.}
    \label{table:issues}
\end{table*}

However, the analysis also revealed that the uneven task distribution, mentioned by students in the survey, was also not fixed by employing Kanban:
The number of unique assignees that worked on user stories did not significantly change and not all developers (19) were assigned a user story.
The more dynamic nature and focus on throughput of Kanban was also apparent in analyses concerning commits. 
While diff sizes of commits in the Kanban week did not differ from those of previous Scrum Sprints (see top of Table~\ref{table:commits}), many more commits overall were made (see bottom of Table~\ref{table:commits}).
Students' call for a more thorough introductions to helpful merge strategies is strengthened by the fact that the mean amount of merge commits almost tripled (from 51.5 to 142) from the last Scrum Sprint to the Kanban week.

\begin{table*}[!htbp]
    \centering
    \begin{tabular}{|r|r|r|r|r|r|r|r|r|r|r|r|r|}
        \hline
         & \multicolumn{3}{c|}{\textbf{Files changed per commit}} & \multicolumn{3}{c|}{\textbf{Insertions per commit}} & \multicolumn{3}{c|}{\textbf{Deletions per commit}} \\ \hline
        \textbf{Iteration} & \textbf{Mean} & \textbf{Median} & \textbf{Stdev} & \textbf{Mean} & \textbf{Median} & \textbf{Stdev} & \textbf{Mean} & \textbf{Median} & \textbf{Stdev} \\ \hline
        Sprint 1 & 8.79 & 4 & 15.5 & 109 & 17 & 282   & 58  & 6 & 200  \\ \hline
        Sprint 2 & 4.33 & 3 & 4.3  & 33  & 14 & 65    & 29  & 8 & 70   \\ \hline
        Sprint 3 & 4.33 & 2 & 7.5  & 139 &  9 & 2004  & 323 & 6 & 5952 \\ \hline
        Sprint 4 & 4.43 & 3 & 6.23 &  32 & 10 & 72    & 26  & 7 & 73   \\ \hline
        Kanban   & 3.67 & 2 & 4.47 &  37 & 10 & 69    & 25  & 6 & 50   \\ \hline
    \end{tabular}

    \vspace{4mm}

    \begin{tabular}{|r|r|r|r|r|r|}
        \hline
        & \multicolumn{5}{|c|}{\textbf{Mean per week}} \\ \hline
        \textbf{Iteration} & \textbf{Non-merge commits} & \textbf{Merge commits} & \textbf{Files changed} & \textbf{Insertions} & \textbf{Deletions} \\ \hline
        Sprint 1 &  46 &   4.5 &   400 &  4948 &   2625 \\ \hline
        Sprint 2 &  74 &  23.5 &   321 &  2437 &   2153 \\ \hline
        Sprint 3 & 124 &  49.3 &   549 & 17580 & 122891 \\ \hline
        Sprint 4 & 138 &  51.5 &   619 &  4435 &   3664 \\ \hline
        Kanban   & 289 & 142.0 &  1069 & 10642 &   7320 \\ \hline
    \end{tabular}
    \caption{Commit statistics by iteration. Normalized by commit (top) and per week (bottom). Some iterations had different lengths.}
    \label{table:commits}
\end{table*}

The collection and evaluation of students' development data allowed another dimension of analysis, in addition to proven surveys.
Development artifacts such as tickets and commits are necessarily produced during regular development activity and represent information on the executed process and possible adaptations to changes in development processes.
This is especially relevant when comparing this tool to surveys, where not every participant of the course fills out the survey. However, almost every participant produces artifacts (and students that do not produce any are also of interest to educators).
By comparing perceptions of students to what actually happened during the course, as read from development data, insights into those areas of the course that need improvement the most can be gained, e.g. the focus on teaching effective merge strategies.
Using development data, supporting surveys, we were able to improve our assumptions on how students adapt their workflow when employing Kanban.




\section{Related Work}
\label{sec:related_work}
Related work for this paper, in addition to research into agile methods and their application (see Section~\ref{sec:scrum}) is found in the areas of research into how these methods can be taught in education as well as the domain of development process assessment.

\subsection{Teaching Agile Capstone Courses}
Many of the challenges in software development that agile approaches address are also found in the domain of student learning:
complexity (introduction of entirely new concepts), under-defined problems (students are not familiar with problem space), time-boxed development with frequent team meetings (semesters and regular lectures), and inevitable change (applying new knowledge).
Scrum can be viewed not only a software development method, but also as a general learning strategy~\cite{Wallace2012}.
Therefore, educating university students on agile processes using capstone courses, where these strategies can be applied first-hand, seems like a natural fit.
In a literature review in 2015, Mahni\v{c} identified 23 primary studies on the topic of teaching Scrum in software engineering courses~\cite{Mahnic2015}.
The author points out that all studies emphasized the need for Scrum to be taught taught using practical projects.
Therefore, the usage of Scrum in capstone projects, requiring students to work in teams, was the most widely adopted strategy, described in seven studies~\cite{Mahnic2015}.
More recent research in the context of agile capstone courses has focused on differences in Scrum usage between high and low performing teams, concluding that while the frequency of Scrum usage did not differ significantly, high performing teams applied Scrum practices more thoroughly~\cite{Paasivaara2017}.

Most capstone courses previously described in literature focus on the Scrum methodology.
A systematic literature review in 2013 by Ahmad et al. on the application of Kanban in software development identified 19 relevant primary studies. Of these, however, none dealt with educational issues~\cite{Ahmad2013}.
The same authors later conducted a study on student perceptions towards the software factory as a learning environment, where they reported the use of Kanban boards~\cite{Ahmad2014}.
Mahni\v{c} et al., having identified the gap in the existing literature on using Kanban in software engineering education, describe a course design similar to the one proposed here.
The authors divided the student teams of their Bologna master's program course into two groups: one using Scrumban and the other Kanban.
Students were encouraged to experiment with Kanban and Scrum practices and to search for improvements in their software development process by focusing on the average lead time~\cite{Mahnic2015a}.



\subsection{Evaluating Development Processes}
There has been an ongoing debate on the compared effectiveness of Scrum or Kanban usage. 
Lei et al., in a study from 2017, note that their literature review found a lack of statistical evidence on the topic~\cite{Lei2017}.
Based on the six project management factors of the Project Management Body of Knowledge (PMBOK 4.0), the authors performed a survey with 35 respondents to evaluate Scrum and Kanban in industry.
While Lei et al. point out that Kanban may be better suited for managing project schedules, they conclude that based on their results, both Scrum and Kanban lead to the development of successful projects~\cite{Lei2017}.

In the domain of education, students' development processes are often evaluated in respect to the learning goals of the course.
Mahni\v{c} notes capstone courses provide an opportunity to study student' perceptions of agile practices.
He states that students behave similarly to young professionals in capstone courses, as these take place at the very end of their studies~\cite{Mahnic2015}.

Development processes in capstone courses are generally evaluated in only a few different ways:
\begin{itemize}
    \item Using surveys on students' perceptions of their executed process and the value of different dimensions involved in the project~\cite{Bastarrica2017}
    \item Employing tutors that guide student teams and collect insights into teams~\cite{Scharf2013}
    \item Taking end of term exams~\cite{Kropp2013}
    \item Collecting and analysing the development artifacts produced during the course~\cite{Matthies2016}
\end{itemize}

\section{Conclusion}
\label{sec:conclusion}
In this paper we introduce a university capstone software engineering course design, teaching agile methodologies and focusing on the hands-on practice of Scrum and Kanban.
By first introducing the more prescriptive Scrum, with clearly defined meetings, roles and artifacts, students are free to experiment and get comfortable in their self-organized teams, while being able to rely on a given structure.
Kanban can then be introduced to ease the inevitable phase of the project where the software is prepared for release while giving students a more thorough understanding of the agile landscape, beyond Scrum. 
Surveys at the end of the course revealed positive attitudes towards the course and the introduction of Kanban. Students stated that they endorsed the idea of using Kanban at the end of the course for a short period of time.
An analysis of development data produced during the course, both during Scrum sprints as well as during Kanban application, corroborated the learning target of students having adapted their process to a new methodology.